\documentclass[prd,twocolumn,showpacs,preprintnumbers,nofootinbib,amsmath,amssymb,floatfix,superscriptaddress]{revtex4}

\usepackage{amsmath}
\usepackage{amssymb}
\usepackage{graphicx}

\newcommand{\beq}{\begin{equation}}
\newcommand{\eeq}{\end{equation}}
\newcommand{\bea}{\begin{eqnarray}}
\newcommand{\eea}{\end{eqnarray}}

\begin{document}

\preprint{BA-TH636, BI-TP2010/37}

\title{Chromoelectric flux tubes in QCD}

\author{Mario Salvatore Cardaci}
\email{salvatore.cardaci@fis.unical.it}
\affiliation{Dipartimento di Fisica dell'Universit\`a della Calabria,
I-87036 Arcavacata di Rende, Cosenza, Italy}

\author{Paolo Cea}
\email{paolo.cea@ba.infn.it}
\affiliation{Dipartimento di Fisica dell'Universit\`a di Bari, I-70126 Bari, 
Italy \\
and INFN - Sezione di Bari, I-70126 Bari, Italy}

\author{Leonardo Cosmai}
\email{leonardo.cosmai@ba.infn.it}
\affiliation{INFN - Sezione di Bari, I-70126 Bari, Italy}

\author{Rossella Falcone}
\email{rfalcone@physik.uni-bielefeld.de}
\affiliation{Fakult\"{a}t f\"{u}r Physik, Universit\"{a}t 
Bielefeld, Postfach 100131, D-33615 Bielefeld, Germany}

\author{Alessandro Papa}
\email{papa@cs.infn.it}
\affiliation{Dipartimento di Fisica dell'Universit\`a della Calabria,
I-87036 Arcavacata di Rende, Cosenza, Italy \\
and INFN - Gruppo collegato di Cosenza, I-87036 Arcavacata di Rende, Cosenza, 
Italy}

\date{\today}% It is always \today, today,

\begin{abstract}
We analyze the distribution of the chromoelectric field generated by  a static quark-antiquark pair
in the SU(3) vacuum and revisit previous results for SU(2). We find that the transverse profile
of the flux tube resembles the dual version of the Abrikosov vortex field distribution.
We give an estimate of the London penetration length of the chromoelectric field in the
confined vacuum. We also speculate on the value of the ratio between the penetration lengths 
for SU(2) and SU(3) gauge theories.
\end{abstract}


\maketitle

\section{Introduction}

Color confinement in Quantum Chromo-Dynamics (QCD) is a long-distance behavior whose understanding continues 
to be a challenge for theoretical physics~\cite{Bander:1980mu,Greensite:2003bk}.
Lattice formulation of gauge theories allows us to investigate
the confinement phenomenon in a non-perturbative framework. In particular, Monte Carlo simulations can produce
samples of vacuum configurations that can be used to get insight into the non-perturbative sector of QCD.
Tube-like structures emerge by analyzing the chromoelectric field between 
static quarks~\cite{Fukugita:1983du,Kiskis:1984ru,Flower:1985gs,Wosiek:1987kx,DiGiacomo:1989yp,DiGiacomo:1990hc,Singh:1993jj,Cea:1992sd,Matsubara:1993nq,Cea:1992vx,Cea:1993pi,Cea:1994ed,Cea:1994aj,Cea:1995zt,Bali:1994de,Haymaker:2005py,D'Alessandro:2006ug}.
Such tube-like structures naturally
lead to linear potential and consequently to a ``phenomenological'' understanding of color confinement.
\par
An intriguing model was conjectured long time ago by 't Hooft~\cite{'tHooft:1976ep} and Mandelstam~\cite{Mandelstam:1974pi} to explain the formation of chromoelectric flux tubes in QCD vacuum.
It relies on the hypothesis that QCD vacuum behaves like a coherent state of color magnetic
monopoles. This amounts to say that the vacuum of QCD is a 
magnetic (dual) superconductor~\cite{Ripka:2003vv}. 
According to this picture the (dual) Meissner effect naturally accounts for  the observed color flux tubes.
There are clear analogies with  the usual superconductivity where, as found by Abrikosov~\cite{Abrikosov:1957aa},  a tubelike structure arises as a solution of Ginzburg-Landau equations.
Nielsen and Olesen also found tubelike or vortex solutions in their study of the Abelian Higgs model~\cite{Nielsen:1973cs}. In particular they showed that a vortex solution exists independently of the fact that vacuum behaves like a type I or type II superconductor.
\par
Even if the dynamical formation of color magnetic monopoles  is not explained by the 't Hooft construction, 
lattice calculations~\cite{Shiba:1994db,Arasaki:1996sm,Cea:2000zr,Cea:2001an,DiGiacomo:1999fa,DiGiacomo:1999fb,Carmona:2001ja,Cea:2004ux,D'Alessandro:2010xg}  have given numerical evidence in favor of their condensation in the QCD vacuum. 
However, as observed in Ref.~\cite{'tHooft:2004th} in
connection with dual superconductivity picture, magnetic monopole condensation in the confinement mode could be the consequence rather
than the origin of the confinement mechanism that  actually could depend on additional dynamical causes. 
\par
No matter whether monopole condensation and dual superconductivity could give an exhaustive account of color confinement, it is worth to analyze
tubelike structure in the QCD vacuum using the ``phenomenological'' frame of dual superconductivity picture. 
In previous studies~\cite{Cea:1992vx,Cea:1993pi,Cea:1994ed,Cea:1994aj,Cea:1995zt} of  SU(2) confining vacuum it was recognized the presence  in lattice configurations of color flux tubes made up by the chromoelectric fields directed along the line joining a static quark-antiquark pair. By adopting the language of the dual superconductivity, the transverse size of the chromoelectric flux tube was interpreted as the London penetration length in the Meissner effect. By measuring the penetration length on lattice gauge configurations in the maximal Abelian gauge and without gauge fixing, it was also shown that the so-called London penetration length is a physical gauge-invariant quantity.
Moreover starting from the simple definition of the string tension as the energy stored into the flux tube per unit length,  
it was possible to compute the string tension from the measured distribution of the chromoelectric field. In this way an estimate of the string tension was obtained in good agreement with the results in the literature.
\par
In the present work we investigate the  formation of chromoelectric flux tubes in the more physical case of  SU(3) gauge theory.
The main aim is to compute the size of the chromoelectric flux tube in QCD.
The method and the numerical results are reported in Section~\ref{numerical}. In Section~\ref{conclusions} we discuss our results and present our conclusions. 

\section{Color fields on the lattice}
\label{numerical}

\begin{figure}[htb]
\includegraphics*[width=0.65\columnwidth,clip]
{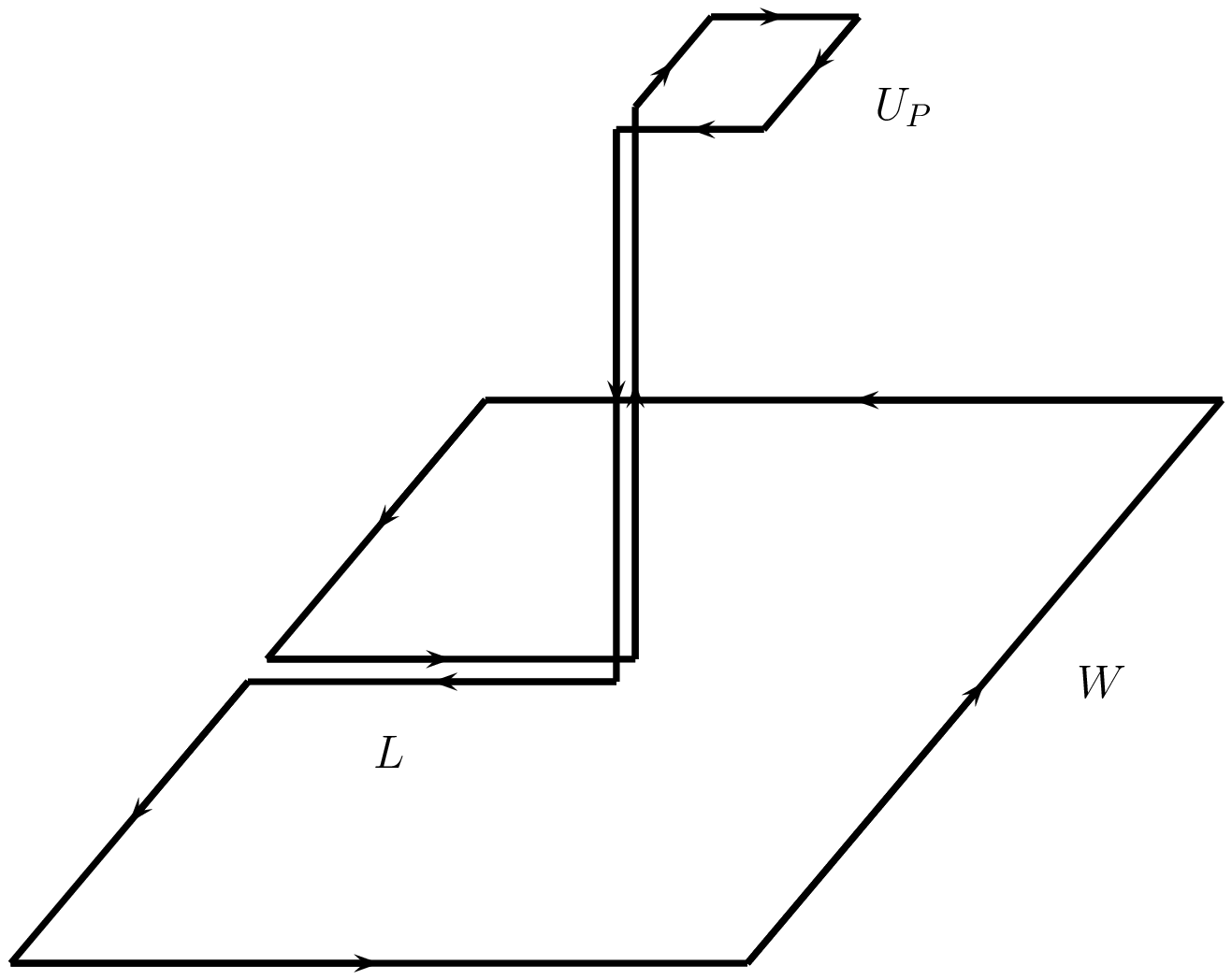} 
\caption{The connected correlator~(\ref{rhoW}) between the
plaquette $U_p$ and the Wilson loop. The subtraction appearing in the
definition of correlator is not explicitly drawn.}
\label{Fig:correlator}
\end{figure}

The field configurations produced by a static quark-antiquark pair in SU(N) gauge theory can be 
explored~\cite{DiGiacomo:1989yp,DiGiacomo:1990hc,Kuzmenko:2000bq,DiGiacomo:2000va} by means of the following 
connected correlation function:
\begin{equation}
\label{rhoW}
\rho_W = \frac{\left\langle {\rm tr}
\left( W L U_P L^{\dagger} \right)  \right\rangle}
              { \left\langle {\rm tr} (W) \right\rangle }
 - \frac{1}{N} \,
\frac{\left\langle {\rm tr} (U_P) {\rm tr} (W)  \right\rangle}
              { \left\langle {\rm tr} (W) \right\rangle } \; ,
\end{equation}
where (see Fig.~\ref{Fig:correlator}) $U_P=U_{\mu\nu}(x)$ is the plaquette in the $(\mu,\nu)$ plane connected
to the Wilson loop $W$  by a Schwinger line L,  $N$ is the number of colors.
The correlation function defined in Eq.~(\ref{rhoW}) measures the field strength.
Indeed in the naive continuum limit~\cite{DiGiacomo:1990hc}:
\begin{equation}
\label{rhoWlimcont}
\rho_W  \stackrel{a \rightarrow 0}{\longrightarrow} a^2 g \left[ \left\langle
F_{\mu\nu}\right\rangle_{q\bar{q}} - \left\langle F_{\mu\nu}
\right\rangle_0 \right]  \;,
\end{equation}
where $\langle\quad\rangle_{q \bar q}$ denotes the average in the presence of 
a static $q \bar q$ pair and $\langle\quad\rangle_0$ the average in the vacuum.
According to Eq.~(\ref{rhoWlimcont}) we define the color field strength
tensor as:
\begin{equation}
\label{fieldstrength}
F_{\mu\nu}(x) = \sqrt\frac{\beta}{2 N} \, \rho_W(x)   \;.
\end{equation}
By varying the distance and the orientation of the plaquette $U_P$
with respect to the Wilson loop $W$, one can probe the color field
distribution of the flux tube. In particular, the case of plaquette
parallel to the Wilson loop corresponds to the component of the 
chromoelectric field longitudinal to the axis defined by the static quarks.

\subsection{SU(2)}
\label{SU2}

\begin{figure}[htb]
\includegraphics*[width=0.65\columnwidth,clip]
{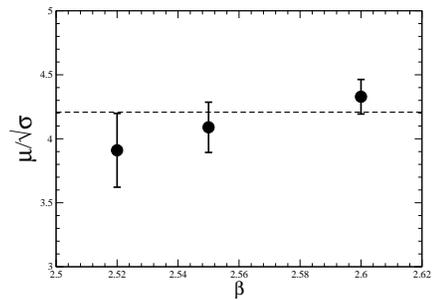} 
\caption{Scaling of the inverse London penetration length with $\sqrt\sigma$
versus $\beta$ in SU(2).}
\label{Fig:sqrtsigma_su2}
\end{figure}

In previous studies~\cite{Cea:1992sd,Cea:1992vx,Cea:1993pi,Cea:1994ed,Cea:1994aj,Cea:1995zt}
the formation of  chromoelectric flux tubes
was investigated in SU(2) lattice gauge theory, both in the maximal Abelian gauge and
without gauge fixing. 

The main result of that works was that the flux tube is almost completely 
formed by the longitudinal chromoelectric field, $E_l$, which is constant 
along the flux and decreases rapidly in the transverse direction $x_t$.

The formation of the chromoelectric flux tube was interpreted as dual Meissner
effect in the
context of the dual superconductor model of confinement. In this context
the transverse shape of the longitudinal chromoelectric field $E_l$ should resemble
the dual version of the Abrikosov vortex field distribution. Hence 
the proposal was advanced~\cite{Cea:1992sd,Cea:1992vx,Cea:1993pi,Cea:1994ed,Cea:1994aj,Cea:1995zt} to fit
the transverse shape of the longitudinal chromoelectric field according to 
\begin{equation}
\label{London}
E_l(x_t) = \frac{\Phi}{2 \pi} \mu^2 K_0(\mu x_t) \;,\;\;\;\;\; x_t > 0\;.
\end{equation}
Here, $K_0$ is the modified Bessel function of order zero, $\Phi$ is
the external flux, and $\lambda=1/\mu$ is the London penetration length. 
Equation~(\ref{London}) is valid if $\lambda \gg \xi$, $\xi$ being the coherence 
length (type-II superconductor), which measures the coherence of the magnetic 
monopole condensate (the dual version of the Cooper condensate). 

Moreover, in Ref.~\cite{Cea:1995zt} it was found that the inverse 
penetration length $\mu$ exhibits approximate scaling with the
string tension $\sigma$, leading to $\mu/\sqrt{\sigma}=4.04(18)$, based 
on a numerical study on lattices $16^4$, $20^4$ and $24^4$ with poor 
statistics (20-100 configurations). Assuming $\sqrt{\sigma}=420$~MeV,
this amounts to have a penetration length $\lambda=0.118(5)$~fm, in 
good agreement with the results obtained in Ref.~\cite{Suzuki:2007jp} on 
a $32^4$ lattice.

In this work, we first repeated the determination of $\mu$ in SU(2) 
with a much larger statistics (details on the numerical setup are 
postponed to the next subsection, where the SU(3) case is considered).
We confirm the scaling of $\mu$ with the string tension $\sigma$ (see Fig.~\ref{Fig:sqrtsigma_su2})
from which we estimate:
\begin{equation}
\label{muSU2}
\mu/\sqrt{\sigma}=4.21(16) \,.
\end{equation}
The result given in the above equation is based on a study on
a  $20^4$  lattice, with a statistics of $1000$ configurations.

\subsection{SU(3)}

The main motivation for repeating the study in SU(3) is to verify the
scaling of $\mu$ with the string tension and to compare
the resulting determination of $\mu/\sqrt{\sigma}$ with SU(2).
This result should provide us with important reference values, that any 
approach aiming at explaining confinement should be able to accommodate.

We performed numerical simulations with the Wilson action and periodic
boundary conditions, using a the Cabibbo-Marinari algorithm~\cite{Cabibbo:1982zn}, 
combined with overrelaxation on SU(2) subgroups. The summary of 
$\beta$ values, lattice size, Wilson
loop size and statistics is given in Table~\ref{Table:runs}. The lattice
size $L$ has been chosen such that the combination $L\sqrt{\sigma}\gtrsim 4$.
The size of the Wilson loop entering the definition of the operator given
in Eq.~(\ref{rhoW}) has been fixed at $L/2-2a$. In order to reduce the 
autocorrelation time, measurements were taken after 10 updatings. The 
error analysis was performed by the jackknife method over bins at different 
blocking levels.

\begin{table}[htb]
\begin{center}
\begin{tabular}{|c|c|c|c|}
\hline
$\beta$ & lattice & Wilson loop & statistics \\
\hline
 5.90   & $18^4$  & $7\times 7$  &  5.k   \\ 
 6.00   & $20^4$  & $8\times 8$  &  4.5k \\ 
 6.05   & $22^4$  & $9\times 9$  &  3.6k \\ 
 6.10   & $24^4$  &$10\times 10$ &  2.4k \\ 
\hline
\end{tabular}
\end{center}
\caption{Summary of the Monte Carlo simulations.}
\label{Table:runs}
\end{table}

In order to reduce the quantum fluctuations we adopted the controlled
cooling algorithm. It is known~\cite{Campostrini:1989ts} that 
by cooling in a smooth way equilibrium configurations, quantum fluctuations 
are reduced by a few order of magnitude, while the string tension survives 
and shows a plateau. We shall show below that the penetration length behaves 
in a similar way. The details of the cooling procedure are described in 
Ref.~\cite{Cea:1995zt} for the case of SU(2). Here we adapted the procedure
to the case of SU(3), by applying successively this algorithm to various 
SU(2) subgroups. The control parameter $\delta$ was fixed at the value 
0.0354, as in Ref.~\cite{Cea:1995zt}.

\begin{figure}[htb]
\centering
\includegraphics*[width=0.65\columnwidth,clip]
{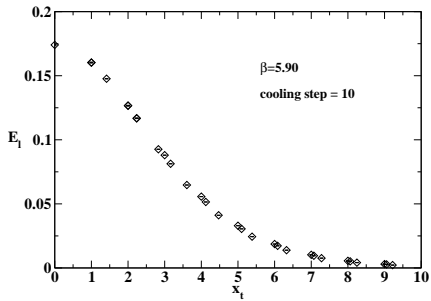} 
\caption{Longitudinal component of the chromoelectric field versus the 
distance $x_t$ at $\beta=5.9$ after 10 cooling steps.}
\label{Fig:El_5.90} 
\end{figure}

A novelty with respect to the study of Ref.~\cite{Cea:1995zt}
is related with the construction of the lattice operator given in
Eq.~(\ref{rhoW}). If the Wilson loop lies on the plane, say, 1-2, then
the Schwinger line can leave the plane 1-2 in the direction, say, 3;
before attaching the plaquette to the Schwinger line, the latter can be 
prolongated further in the direction 4, by one or two links. In this way,
by varying the length of the Schwinger line in the direction 3, one can 
obtain a large set of distances $x_t/a$ between the center of the plaquette 
and the center of the Wilson loop, both integer and non-integer. On each 
configuration we averaged over all possible directions for the relative 
orientation of the Wilson loop to the Schwinger line.

The general strategy underlying this work is the following:
\begin{enumerate}
\item for each $\beta$ we generate an ensemble of thermalized configurations
and, correspondingly, ensembles of ``cooled'' configurations after a number
of cooling steps ranging from 5 to 16;
\item for different values of the distance $x_t$, the longitudinal component 
of the chromoelectric field, averaged over each cooled ensemble of 
configurations, is then determined by means of the operator~(\ref{rhoW}),
with the help of Eq.~(\ref{fieldstrength}) (see, for example, 
Fig.~\ref{Fig:El_5.90}, which shows $E_l(x_t)$ averaged over the ensemble at 
$\beta=5.90$ after 10 cooling steps);
\item for each cooling step, data for $E_l(x_t)$ are fitted with the function
given in Eq.~(\ref{London}) and the parameters $\mu$ and $\Phi$ are extracted;
\item a plateau is then searched in the plot for $\mu$ and $\Phi$ versus
the cooling step.
\end{enumerate}

In Tables~\ref{Table:mu_fit} and~\ref{Table:phi_fit} we report the
results of the fit at the four $\beta$ values considered in this work
for one selected cooling step. When the fit is done on all available data 
for $E_x(x_t)$, above a certain $x_{t,{\rm{min}}}$, the $\chi^2$/d.o.f. is 
very high, thus reflecting the wiggling of data due to the inclusion of 
non-integer distances $x_t/a$. When the fit is restricted to integer values 
of $x_t/a$, the $\chi^2$/d.o.f. turns out to be very reasonable. Remarkably, 
the resulting parameters obtained with the two fitting procedures agree very 
well.

\begin{table}[htb]
\begin{center}
\begin{tabular}{|c|c|c|c|c|c|}
\hline
$\beta$ & cooling & $a\mu$ & $\chi^2$/d.o.f. & $x_{t,\rm{min}}/a$ & data set \\
        &  step   &        &                 &                    &          \\
\hline
 5.90   & 10 & 0.5577(12) & 626. & 6 & all data \\
 6.00   &  9 & 0.51015(92)& 383. & 6 & all data \\
 6.05   & 10 & 0.4730(13) & 133. & 7 & all data \\
 6.10   & 10 & 0.4357(20) &  27. & 7 & all data \\
\hline
 5.90   & 10 & 0.5557(40) & 1.22 & 7 & integer $x_t/a$ \\
 6.00   &  9 & 0.5099(28) & 2.56 & 9 & integer $x_t/a$ \\
 6.05   & 10 & 0.4735(39) & 1.08 & 8 & integer $x_t/a$ \\
 6.10   & 10 & 0.4349(56) & 0.25 & 8 & integer $x_t/a$ \\
\hline
\end{tabular}
\end{center}
\caption{Summary of the fit values for $a\mu$.} 
\label{Table:mu_fit}
\end{table}

\begin{table}[htb]
\begin{center}
\begin{tabular}{|c|c|c|c|c|c|}
\hline
$\beta$ & cooling & $\Phi$ & $\chi^2$/d.o.f. & $x_{t,\rm{min}}/a$ & data set \\
        &  step   &        &                 &                    &          \\
\hline
 5.90   & 10 & 12.784(57) & 626. & 6 & all data \\
 6.00   &  9 & 11.354(41) & 383. & 6 & all data \\
 6.05   & 10 & 14.40(19)  &  87. & 8 & all data \\
 6.10   & 10 & 12.38(11)  &  27. & 7 & all data \\
\hline
 5.90   & 10 & 13.52(25)  & 1.22 & 7 & integer $x_t/a$ \\
 6.00   &  9 & 12.04(16)  & 2.56 & 7 & integer $x_t/a$ \\
 6.05   & 10 & 14.08(30)  & 1.08 & 8 & integer $x_t/a$ \\
 6.10   & 10 & 12.90(38)  & 0.25 & 8 & integer $x_t/a$ \\
\hline
\end{tabular}
\end{center}
\caption{Summary of the fit values for $\Phi$.} 
\label{Table:phi_fit}
\end{table}

\begin{figure}[htb]
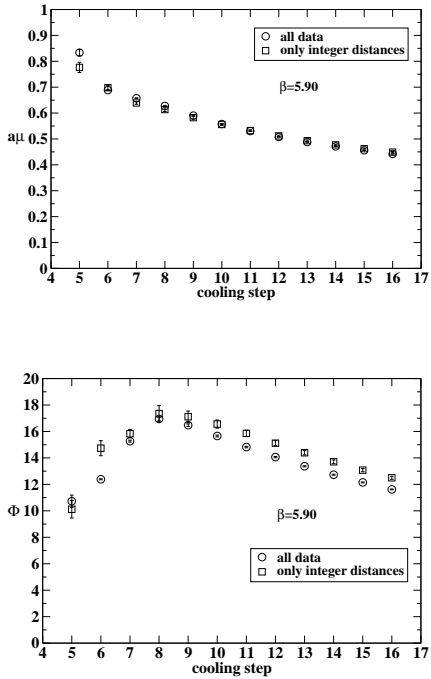

\centering
\includegraphics[width=0.65\columnwidth,clip]{mu_vs_cooling_5.90.eps} 

\vspace{0.9cm}

\includegraphics[width=0.65\columnwidth,clip]{phi_vs_cooling_5.90.eps} 
\caption{(Top) The inverse of the penetration length $a\mu$ at $\beta=5.90$
versus the cooling step. Data are obtained by fitting the transverse profile 
of the longitudinal chromoelectric field with the function~(\ref{London}); 
circles correspond to fit to all available data of $E_l(x_t)$ starting from 
a certain $x_{t,{\rm min}}$, while squares correspond to fit of $E_l(x_t)$ 
for integer values of $x_t/a$.\\
(Bottom) The same for the amplitude of the longitudinal chromoelectric field
$\Phi$.}
\label{Fig:mu_phi_vs_cooling_5.90} 
\end{figure}

\begin{figure}[htb]
\centering
\includegraphics[width=0.65\columnwidth,clip]{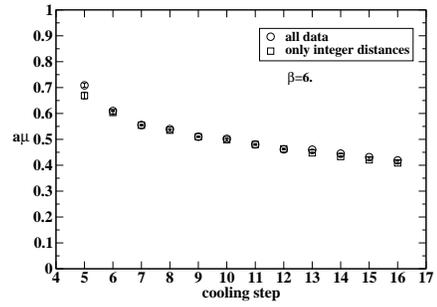} 

\vspace{0.9cm}

\includegraphics[width=0.65\columnwidth,clip]{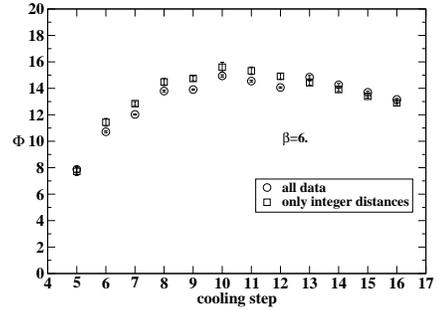} 
\caption{The same as Fig.~\ref{Fig:mu_phi_vs_cooling_5.90} at $\beta=6$.}
\label{Fig:mu_phi_vs_cooling_6.00} 
\end{figure}

\begin{figure}[htb]
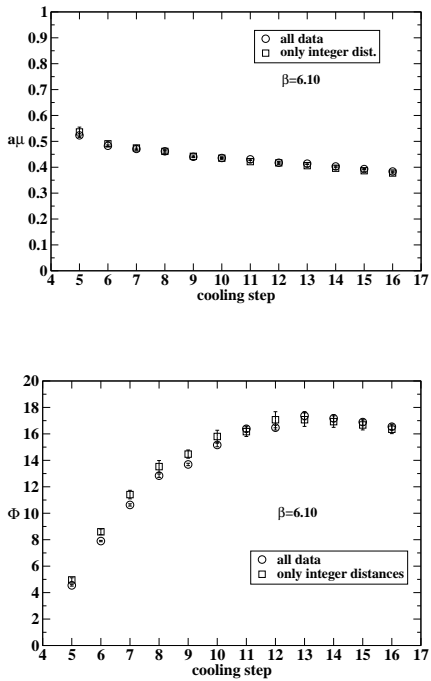

\centering
\includegraphics[width=0.65\columnwidth,clip]{mu_vs_cooling_6.10.eps} 

\vspace{0.9cm}

\includegraphics[width=0.65\columnwidth,clip]{phi_vs_cooling_6.10.eps} 
\caption{The same as Fig.~\ref{Fig:mu_phi_vs_cooling_5.90} at $\beta=6.10$.}
\label{Fig:mu_phi_vs_cooling_6.10} 
\end{figure}

In Figs.~\ref{Fig:mu_phi_vs_cooling_5.90},
\ref{Fig:mu_phi_vs_cooling_6.00}, \ref{Fig:mu_phi_vs_cooling_6.05}, 
\ref{Fig:mu_phi_vs_cooling_6.10}, we show the behavior of $a\mu$ and 
$\Phi$ with the cooling step at the four $\beta$ values considered.
A short plateau is always visible, except for the case of $\mu$ at 
$\beta=5.90$. We take as ``plateau'' value for $\mu$ the value
corresponding to the number of cooling steps given in the second column
of Table~\ref{Table:mu_fit}.

\begin{figure}[htb]
\centering
\includegraphics[width=0.65\columnwidth,clip]{mu_vs_cooling_6.05.eps} 

\vspace{0.9cm}

\includegraphics[width=0.65\columnwidth,clip]{phi_vs_cooling_6.05.eps} 
\caption{The same as Fig.~\ref{Fig:mu_phi_vs_cooling_5.90} at $\beta=6.05$.}
\label{Fig:mu_phi_vs_cooling_6.05} 
\end{figure}

Finally, we studied the scaling of the ``plateau'' values of $a\mu$ 
with the string tension. For this purpose, we have expressed these 
values of $a\mu$ in units of $\sqrt\sigma$, using the parameterization 
\bea
\label{sigma}
a\sqrt{\sigma}(g)&=&f_{SU(3)}(g^2)[1+0.2731\,\hat{a}^2(g) \\
&-&0.01545\,\hat{a}^4(g) +0.01975\,\hat{a}^6(g)]/0.01364 \;, \nonumber
\eea
\[
\hat{a}(g) = \frac{f_{SU(3)}(g^2)}{f_{SU(3)}(g^2(\beta=6))} \;, \;
\beta=\frac{6}{g^2} \,, \;\;\; 5.6 \leq \beta \leq 6.5\;,
\]
\beq
\label{fsun}
f_{SU(3)}(g^2) = \left( {b_0 g^2}\right)^{- b_1/2b_0^2} 
\, \exp \left( - \frac{1}{2 b_0 g^2}\right) \,,
\eeq
\[
b_0=\frac{11}{(4\pi)^2}\;, \;\; b_1=\frac{102}{(4\pi)^4}\;,
\]
given in Ref.~\cite{Edwards:1998xf}.

\begin{figure}[htb]
\centering
\includegraphics[width=0.65\columnwidth,clip]{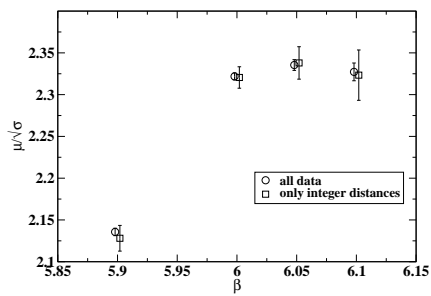} 
\caption{Scaling of the inverse London penetration length with $\sqrt\sigma$
versus $\beta$. Data have been slightly shifted on the horizontal axis for 
the sake of readability.}
\label{Fig:scaling} 
\end{figure}

Figure~\ref{Fig:scaling} suggests that the ratio $\mu/\sqrt\sigma$ displays 
a nice plateau in $\beta$, as soon as $\beta$ is larger than 6.
The scaling of $\mu$ is a natural consequence of the fact that the penetration
length is a physical quantity related to the size $D$ of the flux
tube~\cite{Cea:1992sd,Cea:1992vx}:
\begin{equation}
\label{fluxtubesize}
D  \simeq  \frac{2}{\mu} \;.
\end{equation}
We get the following estimate for the penetration length in SU(3) gauge theory,
\begin{equation}
\label{muSU3}
\frac{\mu}{\sqrt{\sigma}} = 2.325 (5) \;,
\end{equation}
which corresponds to
\begin{equation}
\label{muSU3GeV}
\mu =  0.977 (2) \,{\text{GeV}}.
\end{equation}
We observe that this value is in nice agreement with the determinations
of Ref.~\cite{Bicudo:2010gv}, obtained by using  correlators of plaquette and 
Wilson loops not connected by the Schwinger line, thus 
leading to the (more noisy) squared chromoelectric and chromomagnetic 
fields.

Before concluding this Section we note that the ratio between the penetration lengths
respectively given in Eq.~(\ref{muSU2}) for the SU(2) gauge theory and in Eq.~(\ref{muSU3}) for the
SU(3) gauge theory is:
\begin{equation}
\label{muSU2overSU3}
\frac{\mu_{\text{SU(2)}}}{\mu_{\text{SU(3)}}} = 1.81 (7) \,.
\end{equation}
This result recalls analogous behavior seen in a different study of SU(2) and SU(3)
vacuum in a constant external chromomagnetic background field~\cite{Cea:2005td}. In Ref.~\cite{Cea:2005td} 
numerical evidence that the deconfinement temperature for SU(2) and SU(3) gauge systems in a constant Abelian chromomagnetic field decreases when the strength of the applied field increases was given. 
Moreover, as discussed in Refs.~\cite{Cea:2001an,Cea:2005td,Cea:2007yv}, above a critical strength  $\sqrt{gH_c}$ of the chromomagnetic external background field the deconfined phase extends to very low temperatures. It was found~\cite{Cea:2005td} 
that the ratio between the critical field strengths for SU(2) and SU(3) gauge theories  is 
\begin{equation}
\label{gHratio}
\frac{\sqrt{gH_c}|_{\text{SU(2)}}}{\sqrt{gH_c}|_{\text{SU(3)}}} = 2.03(17) \,,
\end{equation}
in remarkable agreement with the ratio between the penetration lengths  for SU(2) and SU(3) (Eq.~(\ref{muSU2overSU3})).
As stressed in the Conclusions of Ref.~\cite{Cea:2005td}, the peculiar dependence of the deconfinement
temperature on the strength of the Abelian chromomagnetic field $gH$ could be
naturally explained if the vacuum behaved as a disordered chromomagnetic condensate
which confines color charges due both to the presence of a mass gap and the absence
of color long range order, such as in the Feynman picture for Yang-Mills theory in  (2+1) dimensions~\cite{Feynman:1981ss}.

The circumstance that  ratio between the SU(2) and SU(3) penetration lengths agrees within errors with the above discussed ratio of the critical chromomagnetic fields, suggests  us that the Feynman picture of the Yang-Mills vacuum could be a useful guide to understand the dynamics of color confinement. 

\section{Conclusions}
\label{conclusions}

In this paper we present a study of the chromoelectric field distribution
between a static quark-antiquark pair in the SU(3) vacuum, after revisiting some old results
for SU(2) gauge theory~\cite{Cea:1995zt}.
By means of the connected correlator given in Eq.~(\ref{rhoW}) we are able to
compute the chromoelectric field that fills the flux tube along the line joining a
quark-antiquark pair. The transverse behavior of the longitudinal chromoelectric field
can be fitted according to the solution of the London equation for superconductors (Eq.~(\ref{London}))
and gives us information on the so-called penetration length (or inverse size of the flux tube).
We find that the ratio between the penetration lengths respectively for SU(2) and SU(3) 
gauge theories is $1.81(7)$ and agrees, within errors, with the ratio of the 
corresponding critical chromomagnetic fields,  which as discussed at the end of previous Section
could be understood within 
the Feynman picture of the Yang-Mills vacuum.

\begin{acknowledgments}
The work of R.F. has been supported in parts by the grants BMBF 06BI9001 
and the EU Integrated Infrastructure Initiative ``Hadron Physics 2''.
\end{acknowledgments}

%\bibliography{qcd}

\end{document}